\documentclass[5p, twocolumn, sort&compress]{elsarticle}
\usepackage{graphicx,amssymb,color}
\usepackage[dvipsnames]{xcolor}
\usepackage[normalem]{ulem}
\usepackage{color}
\begin{document}


\title{Clustering of floaters on the free surface of a turbulent flow: an experimental study}

\author[cea,uchile]{Pablo Guti\'errez}
    \ead{pagutier@gmail.com}

\author[cea,lyon]{S\'ebastien Auma\^\i tre\corref{cor1}}
\ead{sebastien.aumaitre@cea.fr}

\cortext[cor1]{Corresponding author}

\address[cea]{Service de Physique de l'Etat Condens\'e, DSM, CEA-Saclay, CNRS UMR 3680, 91191 Gif-sur-Yvette, France}
\address[uchile]{Departamento de F\'isica, FCFM, Universidad de Chile, Casilla 487-3, Santiago, Chile} 
\address[lyon]{Laboratoire de Physique, ENS de Lyon, CNRS UMR 5672, 46 all\'ee d'Italie, F69007 Lyon, France}


\begin{abstract}
We present an experimental study of the statistical properties of millimeter--size spheres floating on the surface of a turbulent flow. The flow is generated in a layer of  liquid metal by an electromagnetic forcing. By using two magnet arrays, we are able to create one highly fluctuating flow and another, more stationary flow. In both cases, we follow the motion of hundreds of particles floating at the deformed interface of the liquid metal. We evidence the clustering of floaters by a statistical study of the local concentration of particles. Some dynamical properties of clusters are exposed. We perform spatial correlations between particle concentration and hydrodynamical quantities linked with inertial effects; with vortical motion, and with horizontal divergence (corresponding to compressibility in the surface). From comparing these correlations, we propose the so--called surface compressibility as the main clustering mechanism in our system. Hence, although floaters are not passive scalar and move on a deformed surface, the scenario is similar to the one reported for passive scalar on an almost flat free surface of a turbulent flow.
\end{abstract}

\maketitle

\section{Introduction}

The motion of tracers in turbulent flows has attracted a lot of attention because of its impact in pollutant dispersion in the atmosphere and in the oceans. For instance, floating garbage concentrate in large litters in the middle of the ocean \cite{MartinezEtAl2009, LawEtAl_2010, MaximenkoEtAl_2012, CozarEtAl_2014}.  It also plays a role in the oceanic ecosystem (plankton mixing); in cloud dynamics and rain formation; or in ocean--atmosphere mixing.

At a fundamental level, it has been conjectured that a passive scalar advected by a turbulent flow has a highly intermittent dynamics, even if the flow itself is not intermittent \cite{Kraichnan}. The dynamics of finite size particles with inertia is even more intricate because it involves memory effects and a particular time scale, the Stokes time \cite{MaxeyRiley,FalkovichI,QureshiEtAl_2007,XuBodenchatz_2008}, which is controlled by the particle size and by the mismatch between the densities of the particle and the fluid. As a consequence, inertial particles do not sample the fluid uniformly, thus the phenomenon of preferential concentration emerges (see, for instance, \cite{SquiresEaton1991,Bourgoin,BourgoinXu2014}).

Because of buoyancy, particles of intermediate density stay on an interface between a heavy fluid (e.g. water) and a light fluid (e.g. air): so, particles behave as \emph{floaters}. Hence they experience a \emph{compressibility effect} \cite{CressmanGoldburg_JSP2003,BoffettaEtAl_2004} induced by the motion of the heavy fluid, with sources of upwelling fluid, and sinks of downwelling fluid: floaters are attracted to fluid sinks, and they are expelled from fluid sources. Clustering has been observed for fictive, point--like, fully passive particles floating on an almost--flat free surface. This has been explained by the compressible nature of the flow of particles. In order to isolate the compressible effect, these works combine experimental measurements of the surface flow and digital tracking of point--like fictive particles \cite{Goldburg,GoldburgPhysicaD, Lovecchio}. To some extent, the present work expand these studies (i) to stronger flows inducing surface deformation and (ii) to more realistic finite size floaters subject to inertia and capillarity.

Floaters are also transported by surface waves. Stokes drift is responsible in the case of traveling waves \cite{HerterichHasselmann1982,VandenBroeck1999,SantamariaEtAl2013}. For standing waves, on the other hand, gentle transport of finite size floaters is observed in small scale experiments, where floaters eventually agglomerate \cite{Falcovich_2005}. It is argued that inertia and surface tension play a fundamental role \cite{Falcovich_2005}, and the covering fraction of the surface is also relevant to trigger more complex collective scenarios \cite{SanliEtAl_2014a, SanliEtAl_2014b}.
Periodic motion in bounded domains can also induce mass transport as a consequence of viscous boundary layers. In acoustics, this phenomenon is well--known as acoustic streaming \cite{Batchelor}, and it appeared to be relevant also for parametrically excited (Faraday) waves \cite{Douady_1990, GordilloMujica_2014}.

The dynamics of particles floating on turbulent flows is thus subjected to several physical constrains, some of them leading to clustering. 
The aim of our study, rather than isolate these mechanisms, is to compare their contributions to clustering, in turbulent flows with a free surface mimicking those in natural contexts (like rivers or oceans). To do so, we pursue three main objectives: (a) to create and describe the flow; (b) to study the dynamics of real nonwetting particles floating on its surface, in particular their ability to form clusters; and (c) to compare the contributions to clustering of the different physical processes involved in our setup. 

To generate the flow, we use a magnetohydrodynamical (MHD) forcing \cite{Bondarenko, Sommeria_1986}. Although it is challenging to perform measurements, this type of forcing has the advantage of producing a turbulent flow with important surface deformation, as it occurs in natural flows. Strong velocity fluctuations and surface deformation come from using a thin layer of liquid metal, of strong electrical currents and of a strong inhomogeneous magnetic field. It could be noticed that such MHD flows are also relevant for industrial applications \cite{Davidson}.
A second advantage of our MHD setup is the ease in changing the geometry of the imposed magnetic field, i. e. controlling the dynamical properties of the flow. While a highly fluctuating flow is created by a regular array of magnets, a random array of magnets generates a much less fluctuating flow. In section \ref{secII}, the details of the experiment and the characteristics of the flow are described, for both magnet arrays.

In section \ref{secIII}, we focus on the dynamics and clustering of floaters. Some mixing properties of floaters are exposed. It appears that above a threshold in the forcing, both magnet arrays exhibit similar mixing properties. Then, clusters are clearly identified by studying the statistical properties of the area of Delaunay triangles linking nearest neighbors. We observe stronger correlations in velocity for particles belonging to a cluster.

In section \ref{secIV}, we evaluate the correlation between the particle concentration and some properties of the flow. The correlations measure the contributions to clustering in our setup. We show that the most probable clustering mechanism comes from the horizontal divergence, linked to a compressible effect for particles at the surface. Effects of curvature and capillarity are also discussed.
Finally, in section \ref{secV} we give the concluding remarks. 

\begin{figure*}
\begin{center}
\includegraphics[width=14cm]{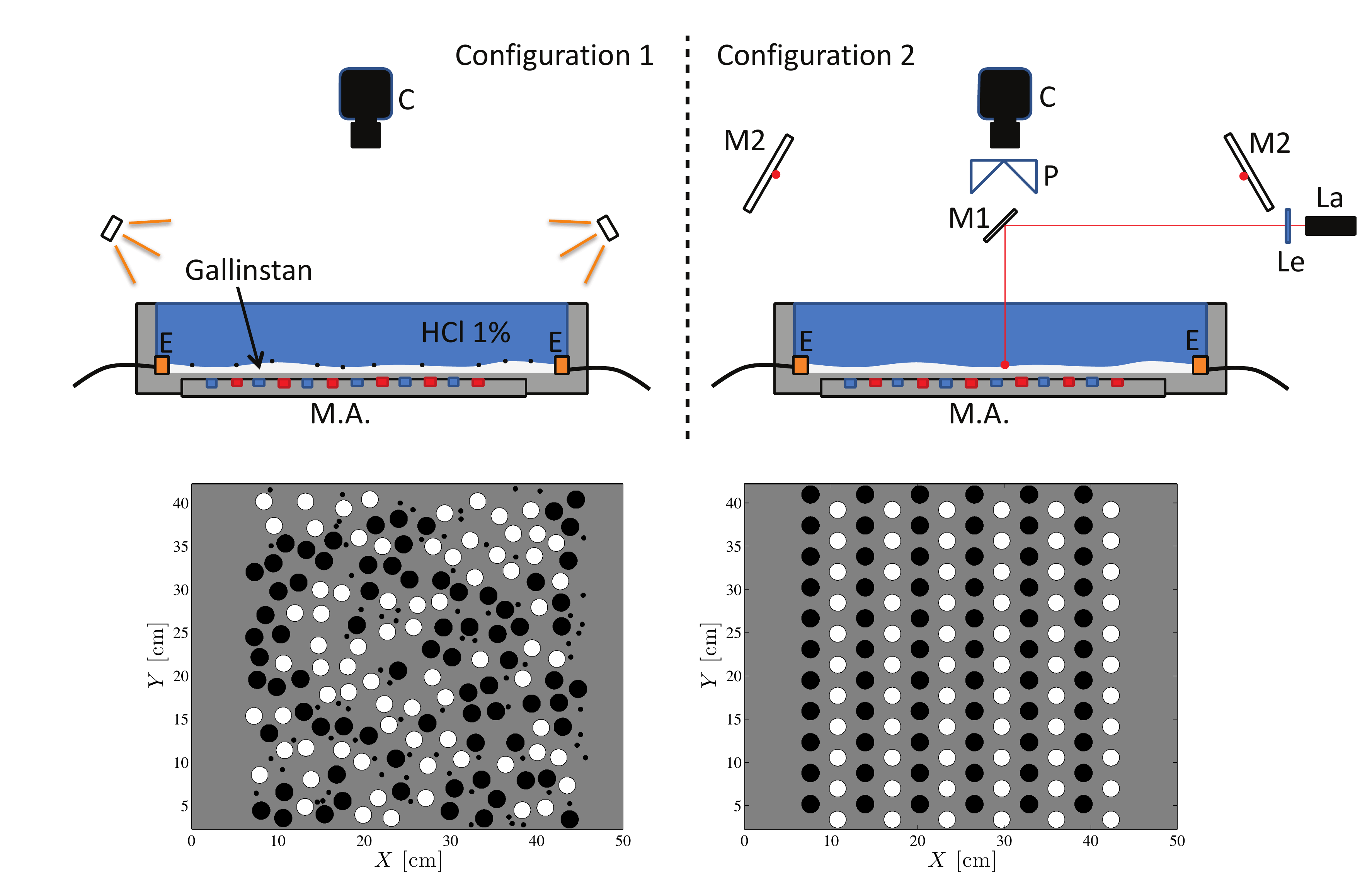}
\caption{(Color online) The experimental device. {\bf Top :} diagram of the experiment and measurement configurations. A 1 cm layer of Gallinstan (GaIn Ti) is placed between two electrodes (E), over a magnet array (M.A.) in a container of $50\times40$ cm$^2$. 
In configuration 1, the camera (C) records the position of particles floating on the Gallinstan. In configuration 2, the beam of the Laser diode (La) is transformed into a laser sheet by the cylindrical lens (Le) and projected on the surface of a mirror (M1). The diffused line, made by the laser sheet on the Gallinstan surface, is tracked with 2 opposite angles by a single Camera (C) by the two mirrors (M2) and the prisms (P). {\bf Bottom :} Sketches of the random and regular magnet arrays used in the experiment. Black and white indicate the polarity of the magnets.}
\label{CellArtI}
\end{center}
\end{figure*}  

\section{Creation and characteristics of a free-surface turbulent flow}\label{secII}

This section concerns the turbulent flow under study. First we present our setup and the measurement techniques. 
Then, we give several orders of magnitude --set by dimensional arguments-- 
that complement the measurements. Finally we present experimental results that reveal the main features of the flow.

\subsection{Setup and methods}

 An electric current, of density $\mathbf{ J} $, in addition to a magnetic field, $\mathbf{ B} $, generate a Lorentz force $\mathbf {F_L} $ inside a conducting body, with $\mathbf{ F_L} = \mathbf{ J }\times \mathbf{ B} $. This force has been used to stir conducting fluids first by Bondarenko {\it et al.} \cite{Bondarenko} in order to induce two--dimensional (2D) turbulence with a well--defined forcing wavelength. To do so, they used an uniform current and magnetic strips with alternating polarity. Later, Sommeria \cite{Sommeria_1986} applied a strong uniform magnetic field and space-dependent distribution of current to generate an almost 2D flow and to study the transition between large--scale structures. Since then, the technique has become a common tool to study 2D turbulence \cite{TabelingEtAl_1991,RiveraWuPRL2000}, instabilities \cite{FauveTabeling,Dauxois, Thess}, chaotic mixing in 2D flows \cite{GollubMixing} and wave--vortex interaction 
\cite{FalconFauve_2009, GutierrezAumaitre_2016}. 
We adopted a similar forcing. However, we used  a layer of liquid metal which allowed us to reach high density currents (up to $1.5\times10^5$ A/m$^2$) with no need of high power nor cooling (the applied voltage is less than 1 V). The use of a horizontal current and vertical dipolar magnets to force the fluid layer, generates horizontal shear, which in turn generates strong vertical vorticity. This strong forcing distorts the interface and induces a vertical velocity component.

 A diagram of the experiment is shown in figure \ref{CellArtI}. It is performed in a plastic (isolating) rectangular container, with a maximal working area of 
$40 \times 50$ cm$^2$.
This container is filled with a layer of Gallinstan up to a depth of $H=1$ cm. Gallinstan is a liquid alloy at room temperature, made of gallium, indium and tin\footnote{From the safety datasheet acc, Guideline 93/112/EC of Germatherm Medical AG,  the Gallinstan is made of 68.5 \% of Gallium, 21.5 \% of indium, 10 \% of Tin. Its kinematic viscosity is $\nu=3.73\times 10^{-7}$ m$^2$/s, its electrical conductivity $\sigma=3.46 \times 10^6$ S/m.}.  
It has a density $\rho=6440$ kg/m$^3$. 
A current up to 600~A is supplied with a \emph{Sorensen DHP Series} Power devise, by two brass electrodes placed along the container walls. Beneath the container, we can choose between two types of inhomogeneous magnetic fields ${\bf B}$, which are produced by two different arrays of strong permanent Neodymium--Iron magnets of 20 mm diameter, as shown in figure \ref{CellArtI}. One array is made with regular lines of alternating polarity and the second one with magnets placed randomly\footnote{The random distribution of magnets was obtained by choosing randomly the coordinates ($x, y$) of each magnet, together with their polarity. However, a balance in polarity is respected.}. Both have a mean distance between magnets of roughly $ l=$ 40 mm. At the bottom of the container, the magnetic field just above each permanent magnet is around 1200 Gauss. The oxidation of the Gallinstan--air interface  creates a thin solid film. To prevent it, the Gallinstan surface is covered by a layer of chlorite acid solution (at concentrations lower than a percent). The acid layer is thick enough (about 10 cm) to make the Gallinstan--acid interface insensitive to the boundary condition at the top of the acid layer. The interfacial tension (or simply the surface tension) was measured\footnote{The interfacial tension was determined in a complementary experiment:  in a smaller container we excited Faraday waves. By measuring simultaneously the wave frequency and wavenumber, the value of the interfacial tension was obtained after fitting the dispersion relation.} to be $\gamma=0.5$~N/m. 

The particles we use along this study are spherical, non--wetting, of diameter $d=1$ mm and of density $ \rho_p \sim 0.3\; \rho$. 
Therefore, they are constrained to stay on the interface, floating on the liquid metal. In order to limit the particle--particle interaction and the particle feedback on the properties of the interface, we put only around $N=$ 200 particles (corresponding to a filling fraction of order of $8\times 10^{-4}$).

\begin{figure*}[t]
\begin{center}
\includegraphics[width=17cm]{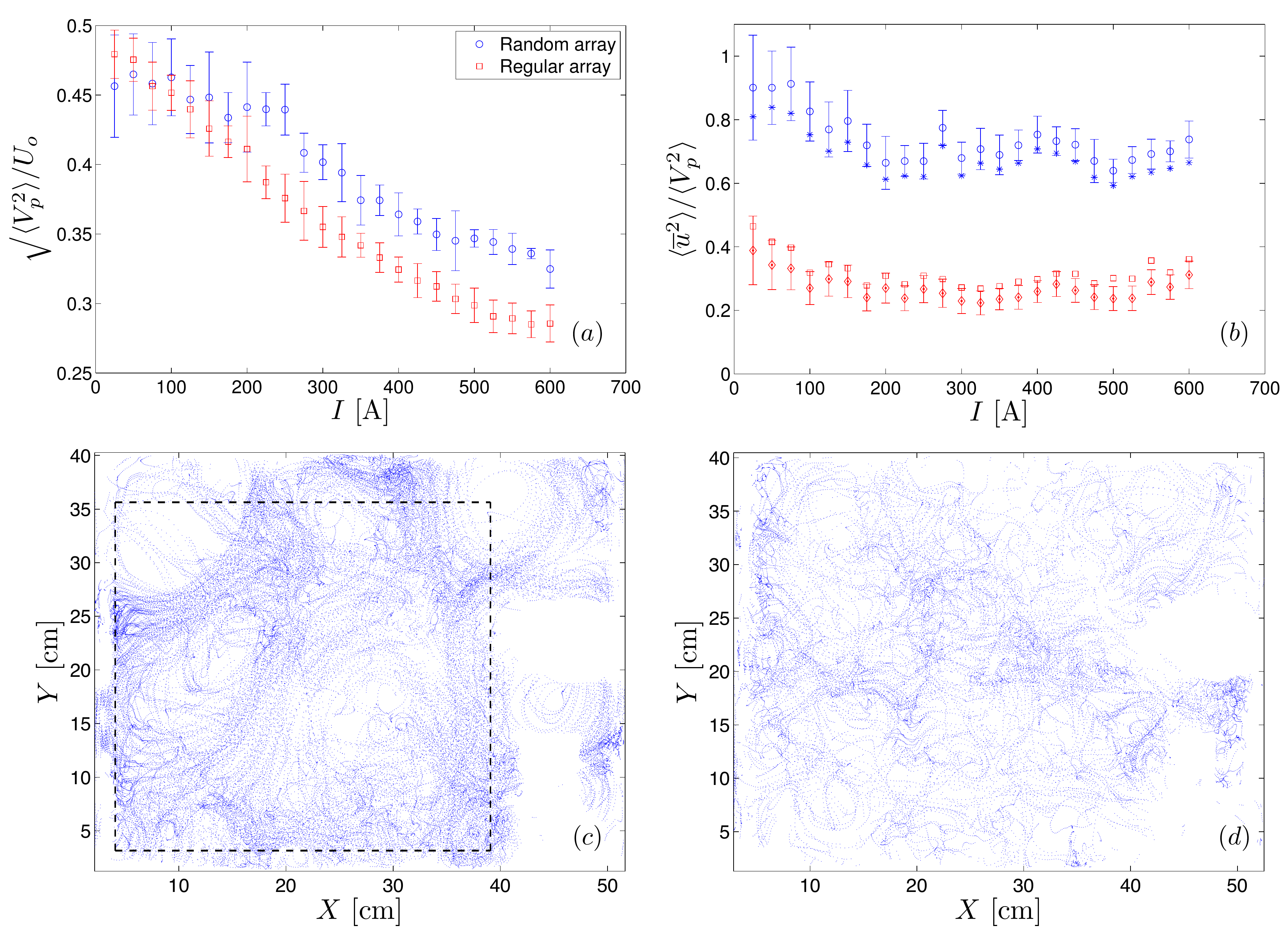}
\caption{(Color online) General features of the flows. (a) shows the evolution of RMS velocities of the floaters normalized by $U_o$, versus the driving intensity $I$. Blue dots correspond to the random magnet array and red squares to the regular magnet array. (b) shows the ratio of the kinetic energy contained in a time--averaged Eulerian flow, divided by the total kinetic energy as a function of $I$. Two spatial resolutions of the Eulerian field are given for each magnets array: the random magnets array with a resolution of $[50\times 50]$  (blue asterisks) and $[100\times 100] $ (blue dots) and the regular magnet array at a resolution of $[50\times 50]$ (red diamonds) and $[100\times 100]$ (red squares). Error bars are estimated from the fluctuations in time and redundant measurements. Figures (c) and (d) show examples of tracked particles for the random and regular arrays of magnets, respectively. In both panels, dots trace the positions of all the particles recorded during 5 s at intervals of 0.1 s, for a forcing current of $I$ = 250~A. See the motion of particles in the supplementary movies, for the random (Movie1-Random) and regular (Movie2-Regular) array of magnets.}
\label{Meanvel}
\end{center}
\end{figure*}

We obtain first the floaters positions. To do so, we use the configuration 1 shown in figure \ref{CellArtI}: a camera of $2000 \times 1700$ pixels$^2$ resolution allows us to detect the 200 particles in the whole container ($d$ = 4 pixels) at a frame rate of 50 Hz. In each image, we obtain the coordinates of the particle centers, despite the difficulties induced by the reflective nature of the liquid metal surface.
From the particles coordinates, we are able both to study the dynamics of particles (see section \ref{secIII}), and to get estimates of the velocity field at the interface \cite{SanliEtAl_2014b}. We access the latter estimates by using standard particle tracking velocimetry (PTV) algorithms. 
The spatial resolution of PTV makes this technique suitable for our flow. In particular, other procedures as particle image velocimetry 
induce spatial average that smooths intense events. The PTV avoids this problem. We compute trajectories using a multi--frame predictive tracking algorithm \cite{OuelletteTrackingLink,OuelletteXuBodenchatz_2006}, which is better suited for fast motion as particle velocity is used to predict the subsequent position (see \cite{OuelletteXuBodenchatz_2006} for comparison with other methods). 
Figure \ref{Meanvel} show representative examples of tracked particles for both the random and regular array of magnets. Supplementary movies (Movie1-Random) and (Movie2-Regular) complement this picture. 
Our PTV technique has an inherent restriction: it gives the velocity of the floaters at the surface instead of the one of the fluid. In other words, finite size floaters act as a filter for very fast or very small velocity fluctuations \cite{QureshiEtAl_2007, XuBodenchatz_2008}. Nevertheless, the obtained velocity do provide general properties of the flow, and consequently, the technique is commonly used in experimental fluid dynamics \cite{CressmanGoldburg_JSP2003, GoldburgPhysicaD, Goldburg, Falcovich_2005,SanliEtAl_2014a,SanliEtAl_2014b,GollubMixing}.

We measure the surface elevation along a line using a classical triangulation technique (see figure \ref{CellArtI}, configuration 2), i. e. tracking the displacement of diffused light spots. This is difficult since the liquid metal interface is poorly diffusive and highly reflective. Indeed, we have to use a very sensitive camera to follow the diffused light, and we had to deal with direct reflective spots suddenly saturating the camera sensor. To recover the information lost due to these spots, we record the line displacement under two opposite angles. Hence, the bright spot in one angle is not seen in the other. Then the whole line displacement can be reconstructed \cite{GutierrezAumaitre_2016}.

\subsection{Dimensionless parameters}
The dimensionless Navier--Stokes Equation, driven by an electromagnetic Lorentz forcing, exhibits a natural velocity scale $U_o=\sqrt{JB  l/\rho}$, which balances the advection term and the Lorentz force. Here we use the forcing length $l$ as the characteristic length of the flow. Thus one gets the Reynolds number $Re=\sqrt{JB/\rho}\cdot~l^{3/2}/\nu$. In our device we can expect $ U_o\sim 30 $ cm/s and $Re~=~3\times10^4$. Such estimates give a Kolmogorov length, $ \eta=\nu^{3/4}/\epsilon^{1/4}\sim 3\times 10^{-2}$ mm with  $\epsilon\sim U_o^3/L$  the energy flux by unit of mass. Note that another choice for the characteristic length, e. g. the container size, would give an unrealistic velocity and Reynolds number. In a thin fluid layer, friction on the bottom plate induces velocity damping. This friction term acts at all scales and induces interaction between structures of different size \cite{Young}. In a liquid metal subject to an electromagnetic forcing, the friction is concentrated at a thin magnetic boundary layer where induction phenomena focus the electric currents and the velocity gradients \cite{Sommeria_1986}. The depth of this layer $e_H=H/H\!a$, is characterized by the Hartmann number $H\!a=\sqrt{\sigma/\rho\nu}BH\leq 45$. 
Hence $e_H$ can be as small as 0.2 mm. One can evaluate this friction strength by the Reynolds number built on Hartmann layer $Re_H=Re_L/H\!a\cdot H/L=\sqrt{JlR/\sigma\nu B}\sim 200$. 
Although the system is highly nonlinear, dynamical features are far from being those of isotropic turbulence or those of 2D turbulence. This is expected since the hypotheses of these two frameworks (isotropy and bidimensionality) are broken in our system. In particular, we observe important deformation of the surface, despite the main component of the vorticity is vertical. 

Gravity and capillarity govern the deformation of the surface depending on the length scale. Gravity dominates at scales larger than the capillary length $l_c=\sqrt{\gamma/\rho g}$. At smaller scales capillarity prevails. In our experiment, $l_c$ corresponds to 3 mm. Two other dimensionless numbers are relevant in free surface flows. One is the Froude number, which is the ratio of the flow velocity over the characteristic velocity of gravity waves, $Fr=U_o/\sqrt{g l}\sim0.2$. As it is not too far from unity, gravity waves generation cannot be completely discarded.
The second number is the Weber number $W\!e=\rho U_o^2 L/\gamma$ ($L$ is a characteristic length). It compares the kinetic energy of the flow and the surface tension energy. At the scale of the forcing $L=l > l_c $ surface tension is negligible ($W\!e\sim 50$). At the floater scale $L=d < l_c$, surface tension cannot be neglected anymore ($W\!e\sim 1$). Thus our millimeter--size floaters are sensitive to capillarity.

\subsection{General features of the observed turbulent flows}\label{IIC}

As we already noticed, floaters velocity give an estimate of the actual flow velocity at the surface, which is useful to infer some general properties of the flow. 
We can first evaluate the root mean square (rms) of particle velocity $\sqrt{\langle V_P^2 \rangle}$. This rms velocity normalized by $U_o$ is shown in figure \ref{Meanvel}--a. The value of $\sqrt{\langle V_P^2 \rangle}$ evolves between 0.5 $U_o$ and 0.3 $U_o$. It is expected that the actual value is smaller than $U_o$ since the estimate is built on the maximum value of the magnetic field. Moreover we measure here the velocity of the floaters that can be smaller than the one of the sustaining fluid. Above a current of 200~A, the ratio decays for both magnet arrays. This may illustrate the fact that, when the forcing is increased, a larger part of the injected power goes to the vertical velocity component, which is excluded from our measurement. Below 200~A, the ratio saturates to a constant value for the random array whereas for the regular one, the ratio decreases continuously. This can be interpreted as a stronger bidimensionalization of the random array at low driving.

To analyze the spatial statistical features of the flow, we arbitrarily define a grid: we divide the container in $N_p = N_w \times N_w$ cells, with $N_w$ typically equal to 50 or 100 (of $0.8 \times 0.8$ cm$^2$ or $0.4 \times 0.4$ cm$^2$, respectively). 
This averaging procedure, used in \cite{SanliEtAl_2014b}, compensates out the inhomogeneity of particles repartition: as the measurement is long enough, all cells are visited by a significant number of particles. Hence, the averaged velocity $\overline{u_{ij}}$ reflects the time averaged velocity field of the surface, $\overline{u}(x,y)$, and it can be used to measure the energy sustained by the mean--flow over the grid. 
On figure \ref {Meanvel}-b we compare the energy $\langle \overline{u}^2\rangle$ of this time--averaged Eulerian velocity field, to the total kinetic energy of the particles. This gives the ratio of the energy contained in the mean--flow.  Almost 80\% of the energy is contained in the mean--flow of the random array, compared to less than 40\% in the case of the regular array. Hence the former is significantly less fluctuating than the latter. We will conveniently take advantage of this difference between the fluctuation properties of both arrays.

The difference between both magnet arrays is also observable on the surface elevation induced by the forcing. We study the spatial variance of the elevation $h(x,y,t)$ along a line perpendicular to the imposed current: $\langle \Delta h^2\rangle(t)=\langle ( h-\langle h\rangle)^2\rangle$, where $\langle~\cdot~\rangle$ stands here for an average along a line.
After time averaging, $h$ seems to follow a power law as a function of the imposed current: $\overline{\langle \Delta h^2\rangle}\propto I^\zeta$, for both magnet arrays. However, the exponent $\zeta$ is different in both cases. It is around 1.6 for the regular array and around 2.2 for the random one.  Moreover, the stationary part of the elevation, $\langle \overline{\Delta h^2}\rangle$, induces around 50\% of the total surface elevation with the random magnet array, whereas it is less than 20\% for the regular array. 
Thus, the results obtained studying the elevation $h$ are consistent with the picture obtained from the averaged velocity field.

\section{Dynamical properties of the floaters}\label{secIII}

We now focus on the dynamics of our finite--size floaters. Despite it is mainly governed by the one of the underlying flow, it may also be influenced by inertia, buoyancy and capillarity. 
We consider first the diffusion and mixing properties. Then, we study the instantaneous spatial distribution of particles. To do so, we focus on the statistical properties of the Delaunay triangles linking the nearest neighbors. The distribution of triangles obtained experimentally is compared with the one obtained from a homogeneous distribution of points. We are able to quantify the level of clustering and to determine a criterion defining clusters from the discrepancy of both distributions. Finally, the properties of the particles velocity inside a cluster are explored.

\subsection{Particles diffusion and mixing}

\begin{figure}
\centerline{\includegraphics[width=8.5cm]{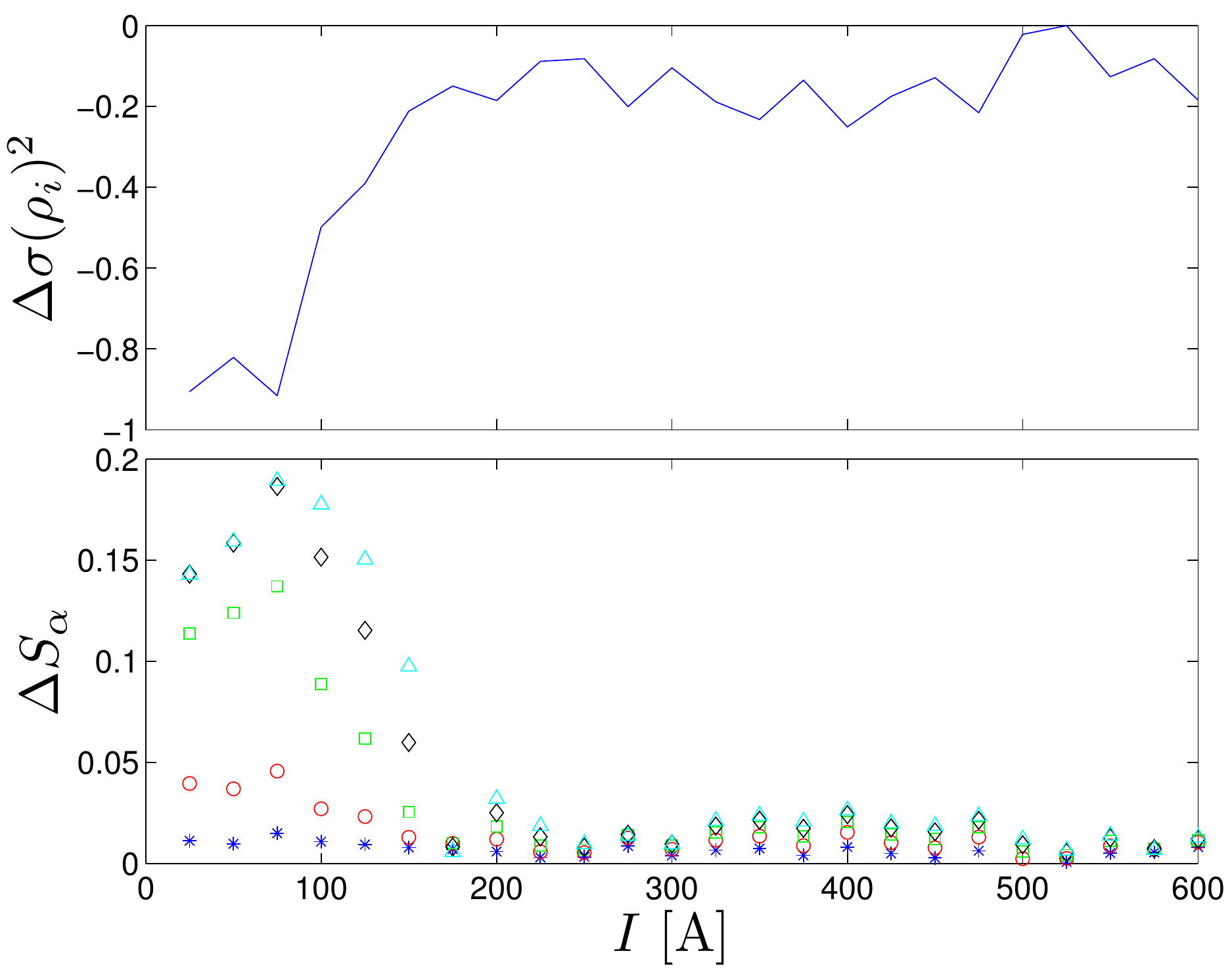}}
\caption{(Color online) Mixing properties as a function of $I$, the applied driving current. {\bf Top:}  Difference between the particles concentration variance of the regular magnets array and the random magnets array, as a function of $I$. {\bf Bottom:} The difference between the R\'enyi entropies $\Delta S_{\alpha}$ of the regular magnet array and the random magnet array, as a function of $I$. $\alpha=1,~\ast$; $\alpha=2,~\circ$; $\alpha=3,~\Box$;   $\alpha=4,~\diamond$; $q=5,~\triangle$. }
\label{Mixing_t}
\end{figure}

\begin{figure*}
\begin{center}
\includegraphics[width=15cm]{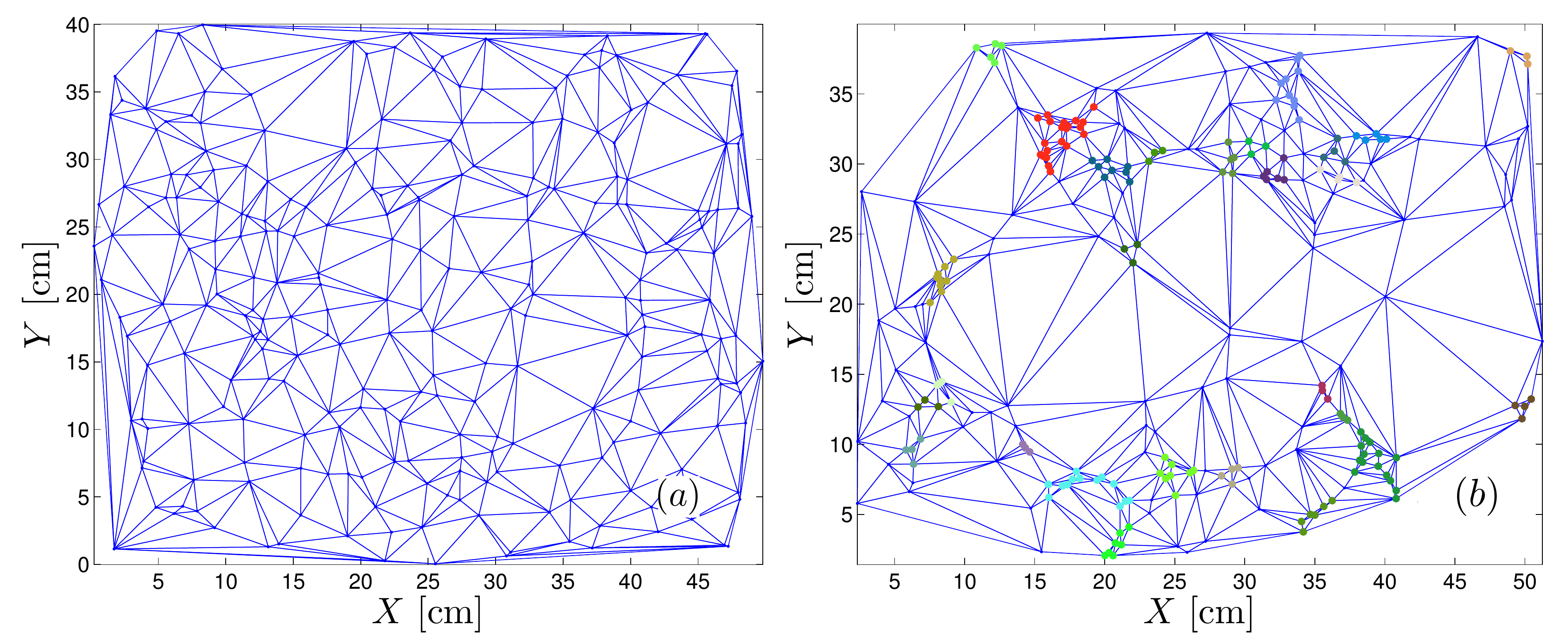}
\caption{(Color online) Position of particles and their corresponding Delaunay tessellation. (a) is for 254 uniformly distributed particles. (b) is for a similar number of particles obtained from a single experimental snapshot, with the random array at $I=$ 300~A. In this last case, color marked points correspond to particles found in different clusters, where triangles areas are smaller than a threshold $A_c = 0.14$ (see text).} 
\label{DelauTri300T100}
\end{center}
\end{figure*}

The usual  way to quantify the diffusion properties of particles is to study the statistical properties of the displacement, $R(t)=\sqrt{(X(t)-X(0))^2+(Y(t)-Y(0))^2}$, for all single particles that we are able to follow during a time $t$. It can be compared with the well-known Brownian motion. Within this framework one has $\langle R(\tau)^2\rangle=D\tau$. The diffusion coefficient $D$ is given by the Einstein formula: $D=2\langle V^2\rangle d^2 /18 \nu,$ where $\nu$ is the fluid viscosity and $d$ is the particle diameter. Unfortunately in our experiment, the range of scales between the forcing and the container size is too small to prevent finite size effects. We observe mainly a ballistic transport until a time $\tau\sim 10\cdot l/\sqrt{\langle V_p^2\rangle}$. It corresponds mostly to the mean time necessary to reach the border after a ballistic flight. All the displacement $\langle R(\tau)^2\rangle$ collapse in a single curve in the ballistic regime if we use $l/\sqrt{\langle V_p^2\rangle}$ as characteristic time unit. Above $\tau\cdot \sqrt{\langle V_p^2\rangle}/l \sim 10$ the displacement curve bends, probably due to finite size effect. The statistical properties of the single--particle displacement are insensitive from the magnet array under study, despite the difference in their fluctuations. 

In order to get a quantitative measurement of the mixing properties of both magnets arrays, we pixelate the container (of total volume $V$) in $N_p$ squares of volume $v_i$, as before. Then we compute the relative concentration in each cell $\rho_i$. It is estimated by counting the number of particles $n_i$ that can be found in each cell $i$ during the experimental run. We normalized it by the mean concentration. Thus, $\rho_i=(n_i/v_i)\cdot(V/N).$ Then we use an usual tool to quantify  mixing: the variance of the relative concentration $\sigma(\rho_i)^2=\langle \rho_i^2-\langle\rho_i\rangle^2\rangle$. The smaller is the $\sigma(\rho_i)^2$, the better the mixing. Other tools to quantify mixing include the relative R\'enyi entropies \cite{GouillardPhD}: \begin{equation}
S_q=\frac{1} {1-\alpha}\log\left(\sum_{i=1} ^{N_p} \rho_i^\alpha\right),$$
\end{equation} which range from 0 to 1. It reaches the limit $S_\alpha=1$ for the perfectly homogeneous mixing. $\alpha=1$ corresponds to the usual Shannon entropy, $\alpha = 2$ is related to the correlation entropy, and higher values of $\alpha$, stress higher fluctuations \cite{GouillardPhD}.

We compute $\sigma(\rho_i)$ and $S_\alpha$  for both magnet arrays (up to $\alpha=5$ for the R\'enyi entropy). Results are shown in figure \ref{Mixing_t}.
The upper panel shows the  difference of the concentration variance between the random array and the regular array, $\Delta\sigma(\rho_i)^2=\sigma_{rg}(\rho_i)^2-\sigma_{rd}(\rho_i)^2$, at various driving current intensities. The bottom panel shows the differences of the R\'enyi entropies  between both arrays at five successive values of $\alpha$. These quantities are estimated during the 60 s of statistically stationary regimes of the experiment. Below $200$~A, there is a discrepancy. It shows that the regular array performs a better mixing. This discrepancy is more important for higher values of $\alpha$. This result underlines that the difference increases when higher fluctuations of the concentration emerge. However  above $200$~A, both magnet arrays have the same mixing properties, despite that the flow produced by the regular array fluctuates more. It should be recalled that a transition around $200$~A has been already observed in the kinetic energy of the particles driven by the random magnet array. 

Therefore, we can conclude that the mixing properties of both flows are equivalent above 200~A, despite their different temporal fluctuations.

\subsection{Clustering characterization}

We are now going to focus on the instantaneous spatial distribution of the floaters. To do so, we use the Delaunay triangles linking three nearest neighbors. We borrow this tool from the study of granular packing \cite{Bonamy,Aste}, and from more recent studies on clustering of inertial particles in fully developed 3D turbulence \cite{Bourgoin}. 
In order to quantify floaters concentration at the surface, we compute the area of the Delaunay triangles. Such triangulations are shown in figure \ref{DelauTri300T100}, for an uniform distribution of 254 points (a) and for the same number of particles tracked on a snapshot of our experiment (b). In solid state physics and granular matter, these tessellations are used to study amorphous states. In the case of a random set of points, the tessellation gives a gamma--distribution $P(\mathcal A)$  for the elementary triangles area $\mathcal A$ \cite{Aste}, with:

\begin{equation}
P(\mathcal A)=\frac{b^a}{\Gamma(a)} \mathcal A^{a-1}\exp(-b\mathcal A),
\label{Gamma}
\end{equation}  

\noindent
$a=\langle\mathcal A \rangle^2/\sigma( \mathcal A)^2$ and $ b=\langle\mathcal A \rangle/\sigma( \mathcal A)^2$. For uncorrelated points uniformly distributed, one expects an exponential distribution with $a=1$ and $ b=1/\langle\mathcal A \rangle$ \cite{pumir}. This is indeed the case in figure \ref{DelauTri300T100}--a, excepting small deviations due to constraints imposed by the container boundaries. This uniform distribution will be used as a reference hereafter. All excess of smaller areas from this reference, can be considered as a trace of clusters of correlated particles. More precisely, an exponent $a<1$ will be the signature of this excess of smaller areas. 
Indeed the probability density function (PDF) of $\mathcal A$ diverges at vanishing values. Hence the most probable value of $\mathcal A$ is 0. The exponential cut--off at large values of $\mathcal A$ is given by the parameter $b$.

\begin{figure}
\centerline{\includegraphics[width=8.5cm]{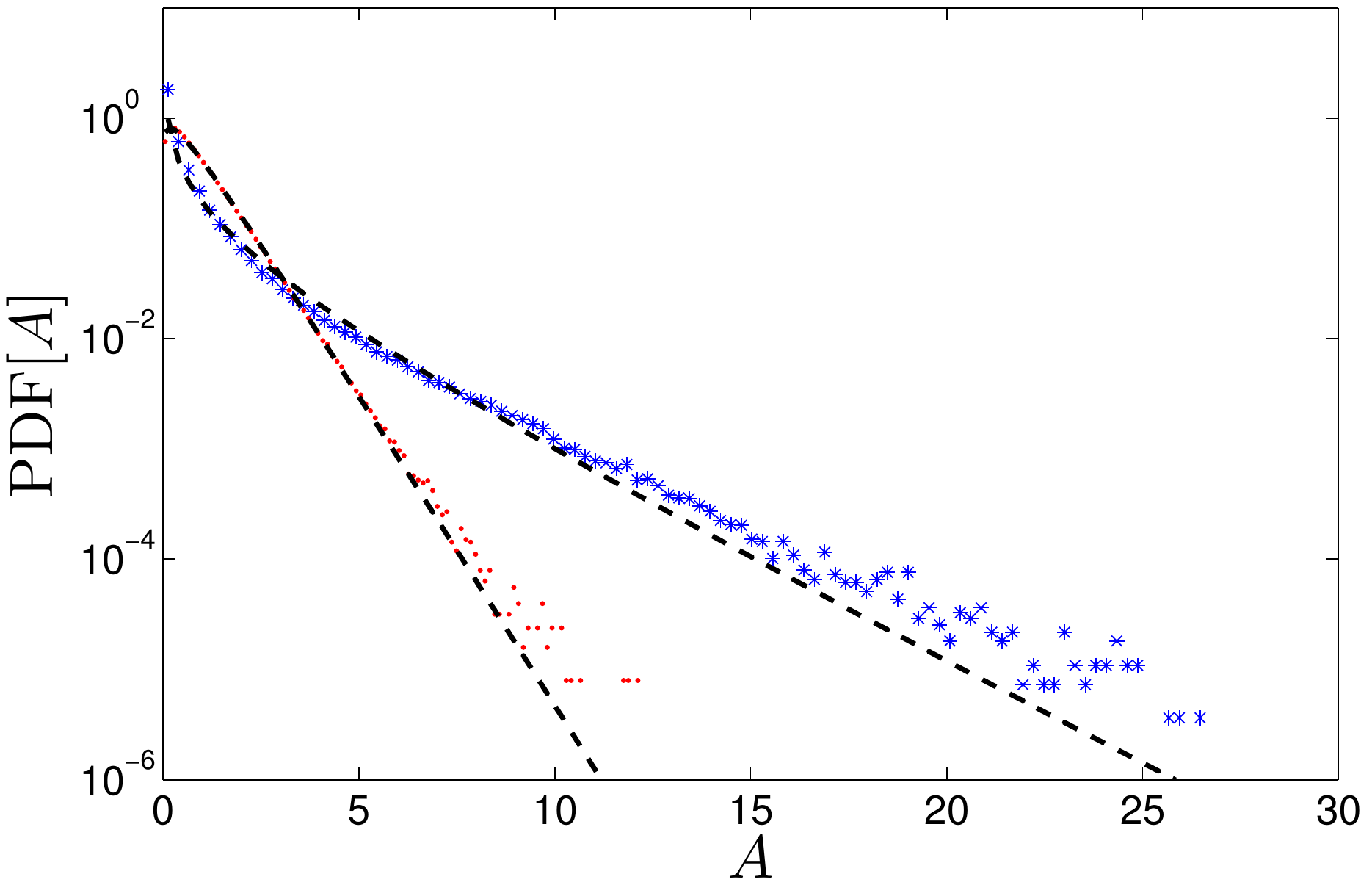}}
\caption{(Color online) Probability density function (PDF) of the normalized Delaunay triangles area. Blue crosses represent the one obtained experimentally, and red dots the one constructed from the uniform distribution of points. Dashed lines correspond to the Gamma distribution of the same average and standard deviation.} 
\label{PDFArd}
\end{figure} 

We study the areas obtained for each snapshot of the experiment. As the number of followed particles and Delaunay triangles can change slightly from time to time, we normalized the area of each triangle, $\mathcal A_i$ by the mean area of triangles at each instantaneous tessellation: $A_i=\mathcal A_i\cdot L_x\cdot L_y/N_t$ where $N_t$ is the number of triangles of the instantaneous tessellation. Hence $\langle A \rangle $=1. Figure \ref{PDFArd} shows the PDF of these normalized areas obtained from 3000 successive snapshots of an experiment performed with the random array and with a driving intensity of $I=300$ A (blue crosses). It also shows the PDF of 3000 realizations of independent successive synthetic tessellations for sets of nearly 200 points uniformly distributed\footnote{For each snapshot of the experiment, we computed an independent set of $n_{ud}$ uniformly distributed points. If the number of particles found in the snapshot varies, $n_{ud}$ varies as well.}. 
For both magnet arrays and all applied currents, the distribution of triangle areas follows  a Gamma distribution (without any fitting parameters once the mean and the standard deviation are given). Note that, if in the case of the synthetic uniform distribution one gets $a$ and $b$ close to one (within 20\% of error due to container boundary), in the case of the experimental PDF one gets $a=0. 311$. This value, smaller than one, is responsible of the cusp observed near 0. The smaller $a$, the stronger is the cusp; i.e. the larger is the excess of smaller areas. Thus, $a$ is indeed a signature of clustering.

Figure \ref{avsI} shows the value of $a$ as a function of the driving current $I$ for both magnet arrays. In both cases, except for the smallest intensity, one gets a decay with $0.24<a<0.5$. There is therefore always a strong clustering. In the case of the regular array, the decay is almost linear, and $a$ goes from 0.5 to 0.35.  For the random array we can observe different behaviors below and above 200~A, once again. Below 200~A, the decay of $a$, going from 0.5 to 0.35 in 150~A, is faster than for the regular array. Above 200~A, the decay rate becomes of the same order for both magnet arrays. 

The areas of the tessellation follow a Gamma distribution (\ref{Gamma}) both for the experimental points and for the synthetic set of uniformly distributed points.  This allows to find an easy criterion to define particles inside a cluster. We consider that a particle is in a cluster if it belongs to a triangle with an area $A$ smaller than a critical value $A_c$. $A_c$ is chosen such that $P_e(A\leq A_c)\geq P_r(A\leq A_c) $, where the indices $r$ and $e$ refer to the synthetic reference distribution and experimental distribution respectively. Using (\ref{Gamma}) and neglecting the exponential decay at large $A$ one gets the following critical value:

\begin{figure}
\centerline{\includegraphics[width=8.5cm]{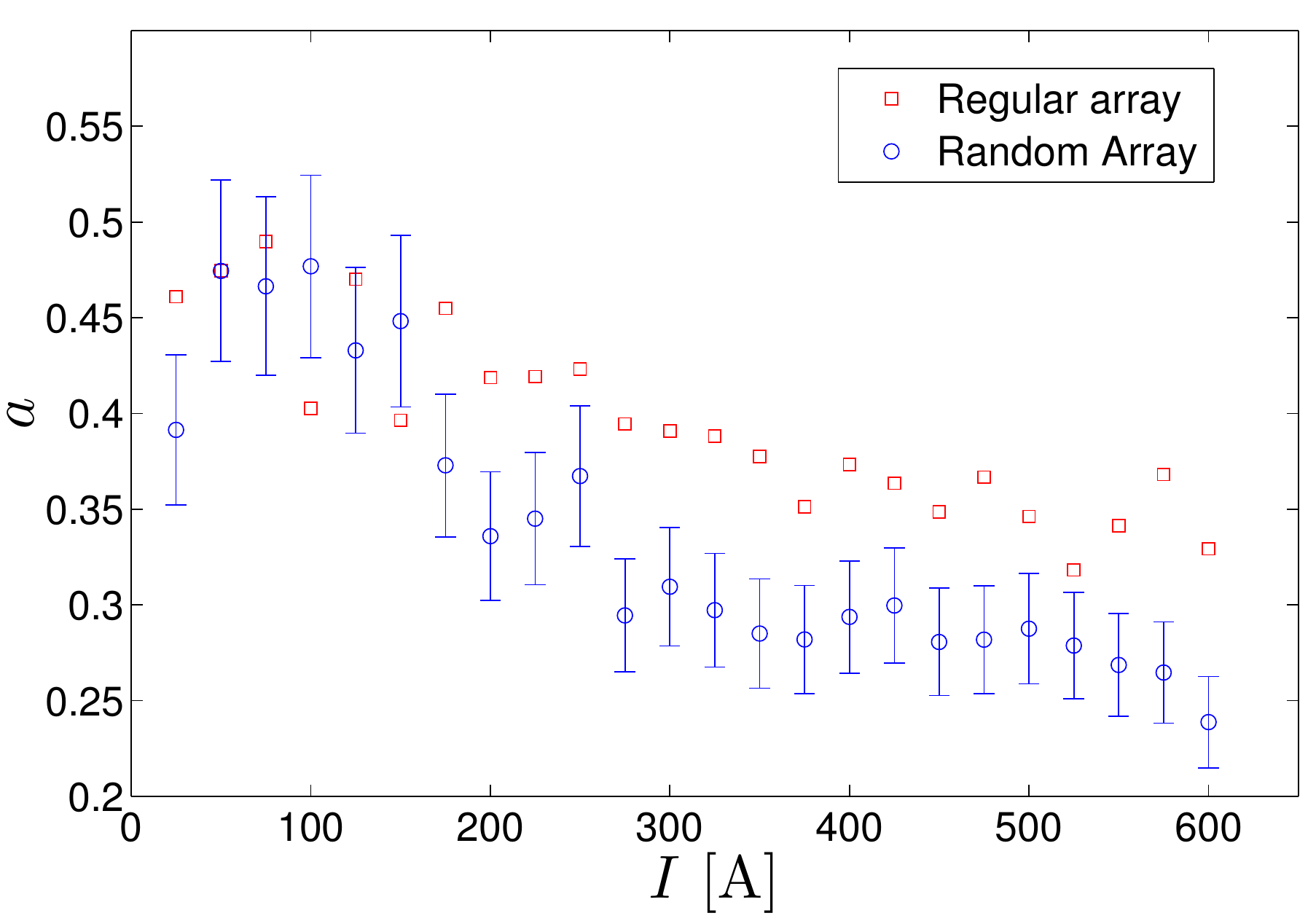}}
\caption{(Color online) Exponent $a$. The values of $a$ are obtained by adjusting the PDF of the Delaunay triangles area with a Gamma distribution. It is done for both magnet arrays, as a function of $I$. Error bars are estimated from the accuracy of the adjustment.} 
\label{avsI}
\end{figure}

\begin{equation}
A_c=\left [\frac{b_r^{a_r}\cdot \Gamma(a_e)}{ b_e^{a_e}\cdot\Gamma(a_r)}\right ]^{\frac{1}{a_e-a_r}}
\label{Ac}
\end{equation}
 
\begin{figure*}
\begin{center}
\includegraphics[width=17cm]{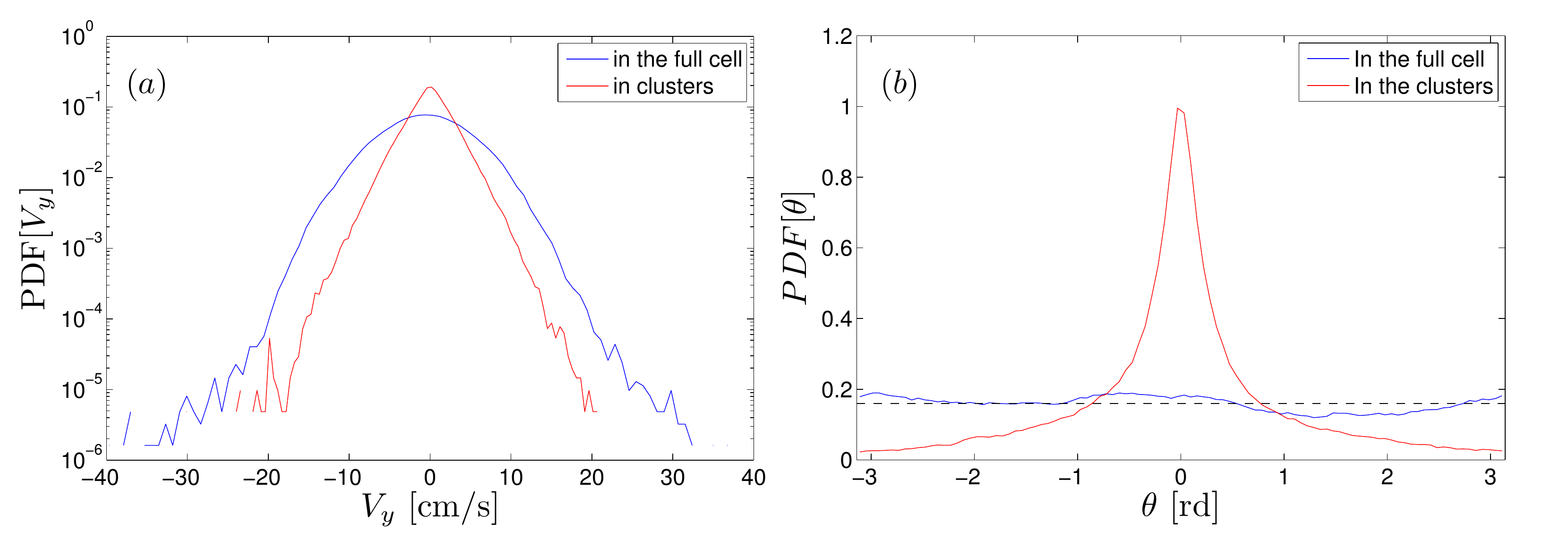}
\caption{(Color online) Statistics of the clusters dynamics. (a)  PDF of the fluctuations around the mean velocity of a cluster for the $y$--component of the velocity of particles belonging to the cluster (red) compared to the unconditional fluctuations (blue).
(b) Angle distribution of the direction of the particles displacement around the direction of the cluster, for particles belonging to a cluster (red). It is compared to the unconditioned distribution of directions (blue) and to the uniform distribution of angles (dashed line).} 
\label{ClusterStat}
\end{center}
\end{figure*}

\noindent
(with $a_r$ and $b_r$ close but not exactly equal to 1). This criterion has been used to define the cluster shown in figure \ref{DelauTri300T100}, in which $A_c=0. 14$. 

We now have the tools to study cluster properties. Indeed, after knowing which particles belong to a cluster, we can compare the velocity fluctuations inside clusters ($A < A_c$) with the whole--sample fluctuations (for any $A$). Figure \ref{ClusterStat}--a exhibits the PDF of the fluctuations of one component of the velocity of particles belonging to clusters $\delta v_y=v_y-V_ {cy} $ around the mean velocity, ${\bf V_c}$, of the cluster. These fluctuations are compared with those of the whole sample. The PDF for particles within the clusters is narrower and its shape, with exponential tails, differs from the nearly Gaussian whole--sample PDF. The flatness, $\langle (\delta v_y-\langle \delta v_y\rangle)^4\rangle/\sigma(\delta v_y) ^4$ of the fluctuations around the cluster velocity is equal to 5.1 whereas the flatness of the whole sample is 3.2, close to the value expected for Gaussian variables. This shows that  the floaters belonging to the same cluster have correlated velocities. The same result can be obtained from the other velocity component. One can also look for correlations in the direction of displacement. We define $\theta_c$ as the angle between the velocity of a floater belonging to a cluster and the velocity of the cluster itself. $\theta_c$ can be compared with the angle distribution of particles velocity in the whole sample. The distribution of $\theta_c$ for clusters is narrower than angle distribution of the whole sample (which is almost uniformly distributed), as shown in figure \ref{ClusterStat}--b. Both results mean that, as expected, the motion inside the clusters are much more coherent that the unconditioned global ones. This strengthens our definition of clusters. 

Finally we check that this coherence is conserved when the forcing is increased.  The ratio of the conditioned over the unconditioned rms velocities is about 0.55 in the regular array whereas it was slightly smaller for the random one (about 0.45). It means that the cluster is slightly more coherent in the second case although the difference is not  significant. In both cases the ratio does not evolve more than 10\% when the current is increased. The rms fluctuations of the angles $\theta_c$ are of the same order for both magnet arrays and the standard deviation is about 40--50\% smaller than the unconditioned case.

\section{Comparing contributions to clustering}\label{secIV}

Now we explore the origin of the clustering. To do so, we correlate the floater concentration with the properties of the flow. As shown before, above 200~A, both magnet arrays have similar mixing and clustering properties (see figures \ref{Mixing_t}--bottom and \ref{avsI}). However, the random array is much less fluctuating, and the time averaged quantities are more representative of the flow properties. We use hereafter only the random array to take advantage of this property. It allows us to correlate in space various quantities averaged in time, for which we have a good spatial resolution. We get the time averaged Eulerian velocity field by the procedure described in section \ref{IIC}. In each cell, we compute the 2D velocity component $\overline{u_i}$ and $\overline{v_i}$ and their derivatives, together with the floaters concentration $\rho_i$, where the index $i$ stands for the $i$-th cell. 

First we check a clustering mechanism similar to the one responsible of the clustering of completely passive tracers at almost--flat surfaces in turbulent flows \cite{GoldburgPhysicaD,Lovecchio,Goldburg}. 
Here, as the fluid is incompressible, one can write the condition $\nabla \cdot {\bf u} = 0$ in terms of a \emph{horizontal divergence} $\nabla_{\bot} \cdot {\bf u_{\bot}} = \partial_x u + \partial_y v = -\partial_z w$.
In a completely two-dimensional fluid the horizontal divergence vanish, whereas it is non-zero in most practical situations \cite{GoldburgPhysicaD,Lovecchio,Goldburg}. The horizontal divergence quantifies the presence of sources or sinks of fluid (as they produce variations in the vertical velocity $w$, even near the free surface where $w$ becomes small). As floaters are constrained to stay at the surface, they cumulate or disperse depending on $\nabla_{\bot} \cdot {\bf u_{\bot}}$, experiencing a compressibility effect. We evaluate the horizontal divergence $-\partial_z w (x, y) |_ {z=h (t)} $ at the surface of the flow. We compute
\begin{eqnarray*} 
q_i =&\partial_z \overline{w_i}(x_i,y_i)|_{z_i=h(t)}\\
=&-\partial_x \overline{u_i}(x_i,y_i)|_{z_i=h(t)} - \partial_y \overline{v_i}(x_i,y_i)|_{z_i=h(t).}
\end{eqnarray*}
Then, we estimate the local correlator $r_i$, between $q_i$ and the normalized time--averaged concentration $\rho_i$, previously introduced: 

\begin{equation}
r_i=\frac{\Delta q_i\cdot \Delta \rho_i}{ \sigma(q_i)\cdot\sigma(\rho_i)},
\label{ri}
\end{equation}
 where $\Delta X$ means $X-\langle X \rangle$. For a driving current of 400~A, the global correlator $\langle r_i\rangle$, averaged over all the cells, is equal to 0.35. The  local correlator (\ref{ri}) is larger than one on 19.8\% of the pixelated area. This underlines spatial coherency between $r_i$ and $ \rho_i$. Although they are not identical, $q_i $ and $\rho_i $ present similar patterns. The correlations are not large, but they are still significant, considering the noise introduced by the coarse--grained gradient and the decoherence induced by time averaging. 

We now compare the previous correlations with those between $\rho_i$ and other hydrodynamic quantities. For instance, the clustering of inertial particles in 3D turbulence is governed by the divergence of the inertial particle velocities, given at first order by $ \nabla \cdot \overline{\bf V_i} \sim-\tau_s \beta_i$ \cite{Maxey_JFM1987,FalkovichI}, with  
\begin{equation}
\beta_i=\bf \nabla\cdot \left(\overline{{\bf u_i}}\cdot \bf \nabla \overline{{\bf u_i}}\right)\label{betai}
\end{equation} and $\tau_s =  d^2(|\rho_p-\rho|)/(18\nu \rho)$, the Stokes time, which takes values near 0.1 s in our conditions.

The spatial distribution of $\beta_i$ allows to define another local correlator 
\begin{equation}
s_i=\frac{\Delta \beta_i\cdot \Delta \rho_i}{ \sigma(\beta_i)\cdot\sigma(\rho_i)}.
\label{si}
\end{equation}
The global correlator $\langle s_i \rangle$ is about  0.07 and the local correlator overcomes unity only on 9\% of the surface. Hence, no common pattern emerges from the comparison between these quantities. Thus, in our setup it is questionable to link the clustering with the inertial effects described, for instance, in \cite{SquiresEaton1991, Bourgoin}.

Particles concentration can also be correlated to the vertical component of the vorticity, which is the strongest component of the vorticity with our forcing. In 3D flows, it was shown that light particles (air bubbles for instance) migrate to zones of intense vorticity, as the pressure is lower there \cite{DouadyCadotVortex1,DouadyCadotVortex2}. Despite the free surface differs from the flow in the bulk, one may expect a similar scenario for floating particles.

From our measurements, the vertical vorticity $\Omega_i$, can be evaluated on the $N_p$ cells. The global correlator with the particle concentration $\langle \Delta\Omega_i\cdot \Delta\rho_i\rangle/(\sigma(\rho_i)\cdot\sigma(\Omega_i))$ gives 0.30, which is comparable, but smaller than $\langle r_i\rangle=0.35$, the global correlation with the horizontal divergence. High correlation between $\Omega_i$ and $\rho_i$ implies that particles concentrate at regions of high vorticity, in opposition with the scenario proposed in \cite{DouadyCadotVortex1,DouadyCadotVortex2}. However, the global correlator, $\langle t_i \rangle = \langle \Delta \Omega_i \cdot \Delta q_i \rangle/(\sigma(q_i)\cdot\sigma(\Omega_i))$ between the vorticity and the horizontal divergence is 0.83, which is large. This strong correlation (much larger than the one between $\Omega_i$ and $\rho_i$) may be explained by secondary flows that are induced in a shallow fluid layer around vertical vortices \cite{Kamp}: upwelling flows merge at the vortex core whereas downwelling flows dive at the vortex edge. 
Therefore, we propose that the apparent correlation between vertical vorticity and the particles concentration is only the result of the correlation between $\Omega_i$ and $q_i$. In a complementary experiment, where secondary flows are absent, light nonwetting floaters move toward the axis of rotation. This experiment, where particles float at the parabolic surface of a fluid in solid--body rotation, will be the subject of a future work. 

To summarize, the hierarchy of correlations we have computed suggest that clustering in our experiment is driven by horizontal divergence. To further discuss these effects, one can introduce a dimensionless factor \cite{Goldburg,BoffettaEtAl_2004} $$C = \frac{\langle (\partial_x u + \partial_y v)^2 \rangle }{\langle (\partial_x u)^2 \rangle + \langle (\partial_y u)^2 \rangle + \langle (\partial_x v)^2 \rangle + \langle (\partial_y v)^2 \rangle}$$ that quantifies the degree of compressibility. It is zero for incompressible 2D flows and it takes values close to 0.5 near the surface, for free surface flows \cite{Goldburg,BoffettaEtAl_2004}. We compute $C$ for the random array of magnets, used to study correlations. It gives a value close to 1/6 for every forcing current $I$ in our experiment, although the values are slightly larger when $I$ is less than 200 A. The value 1/6 is low compared with those obtained previously near the surface \cite{Goldburg, LovecchioEtAl_2015}. It is not still clear for us if the coarse-graining process used to get spatially resolved fields reduces the $C$ coefficient. Another explanation comes from the fact that we are only measuring a projection of the three dimensional deformation of the free surface (i.e. neglecting the vertical velocity). We can sustain this idea by interpreting the 1/6 as the $C$ of the 2D projection of a 3D (homogeneous and isotropic turbulent) flow \cite{BoffettaEtAl_2004}. Our measurement giving $C < 0.5$ does not allow us to see any of the extreme events suggested in \cite{LovecchioEtAl_2015}.

Apart from the aforementioned physical mechanisms, our particles are sensitive to the surface tension because they are smaller than the capillarity length $l_c$. Capillarity makes attractive particles of similar wetting \cite{MahadevanAmJPhys}. The attractive capillarity force between particles decays exponentially with the distance \cite{berhanu}, and thus, it is significant on a characteristic length of order $l_c$. This interaction length is an order of magnitude smaller than the mean free path between floaters in our experiment. Hence, due to the low filling fraction of particles, we expect that capillarity will be initially inefficient to agglomerate floaters. 
However, once the clusters are formed, the attraction could play a stabilizing role and could affect the cluster cohesion. Capillarity force also makes particles sensitive to the local curvature \cite{Falcovich_2005}. However, our experiments with particles floating on the parabolic surface of a rotating fluid (where capillarity seems to play a crucial role), show that the motion is much slower than the one observed in the present experiment. 
A systematic study of the size and wetting properties of floating objects is a very important one \cite{Falcovich_2005, SanliEtAl_2014a,SanliEtAl_2014b,MahadevanAmJPhys,berhanu,QureshiEtAl_2007,XuBodenchatz_2008}, although it is out of the scope of this work\footnote{Although all the results presented here correspond to particles with a diameter of 1~mm, we also used larger particles (3~mm) to verify the reproducibility of the velocity measurements.}. 

A last issue is how spatial inhomogeneity of the floaters reflects the intermittent properties of the underlying flow. 
The high intermittency of the passive scalar stretched and folded by the velocity gradient is revealed by the anomalous scaling of the structure function of the concentration field \cite{Kraichnan}. It has been related to the ramp and cliff inhomogeneous structure of the passive scalar concentration \cite{vergassola}. We do not reach enough resolution to compute the structure function of the concentration field of floaters in our experiment. However, it has been shown that some properties of turbulent flow, or others complex stretched flows, are enclosed in the time evolution of the shape of triangles in 2D (or tetrads in 3D) delimited by Lagrangian points passively advected by the flow \cite{pumir,pumirPRL,ouellettePRL}. Therefore a forthcoming work will be devoted to the study of the time evolution of such distorted triangles, delimited by floaters, in order to underline discrepancies with the passive scalar case \cite{pumir,ouellettePRL}. Moreover, with the tool introduced to define particles belonging to a cluster, we should be able to study the triangle distortion evolution in relation to the particles ability to enter or escape from the clusters and thus to relate spatial inhomogeneity and intermittent properties.

\section{Conclusions}\label{secV}

In the first part of this article, we presented an experimental setup allowing us to generate strongly fluctuating free surface flows of liquid metal, thanks to an electromagnetic forcing.
We used two kinds of magnet arrays, one regular and another random. We apply to them an electrical current $I$ going from 25~A to 600~A. The two magnet arrays have a different level of stationarity. Indeed, the ratio between the energy contained in temporal fluctuations and in the time averaged mean flow, is much larger with the regular array. This result is revealed both by velocity--field estimates and by the surface deformation. 
The random array, which produces a larger mean flow, exhibit several clues of a transition around 200~A: changes appear in the kinetic energy of particles, in the characteristic correlation time and in mixing properties. However, beside fluctuation properties of flows, both magnet arrays behave similarly above this value.

After presenting the flows, we focused on the dynamics of floaters. We shown, in particular, that they do not mix uniformly but tend to form clusters, independent of the magnet array. This was identified by the study of the distribution of the Delaunay triangles areas. The statistics of the area of the triangle linking neighbor particles, follow Gamma distributions. These distributions are singular near zero, illustrating the tendency to form clusters, where the areas are very small. 
By comparing the triangles--area singularity with an uniform reference, a criterion defining clustered particles is obtained. Particles belonging to the same cluster have a coherent displacement. 

To understand the main clustering mechanism, we study the correlations between surface concentration of the floaters and hydrodynamical quantities. One is linked with inertial effects; other with vortical motion; and other with horizontal divergence, that corresponds to compressibility in the surface. 
Knowing that both magnet arrays have similar mixing and clustering behaviors, this analysis is performed with the random array. By doing so, we benefit from stationarity, allowing us to consider time averaged quantities.
Correlations suggests that the main clustering mechanism comes from the horizontal divergence in the surface, which induces a compressible effect in the floaters: they are expelled from the upwelling secondary flow at the vortex core and stretched at the vortex edge.

\section{Acknowledgments}
We would like to thank V. Padilla for building the setup, C. Wiertel--Gasquet for helping us with the automation of the experiment, R. Candelier for the algorithm used to define clusters, F. Daviaud, M. Bonetti and G. Zalczer for helpful discussions. We are indebted to A. Heyries and D. Heyries for their work on the first stage of the experiment. We thanks M. Hjeltman, C. Arratia and L. Gordillo for their careful reading of the manuscript. This work is supported by the ANR Turbulon. PG also received support from the Triangle de la Physique and CONICYT/FONDECYT postdoctorado N$^o$ 3140550.


\bibliographystyle{elsarticle-num}
\bibliography{GutierrezAumaitreEJMflu-BiblioShort}

\begin{thebibliography}{10}
\expandafter\ifx\csname url\endcsname\relax
  \def\url#1{\texttt{#1}}\fi
\expandafter\ifx\csname urlprefix\endcsname\relax\def\urlprefix{URL }\fi
\expandafter\ifx\csname href\endcsname\relax
  \def\href#1#2{#2} \def\path#1{#1}\fi

\bibitem{MartinezEtAl2009}
E.~Martinez, K.~Maamaatuaiahutapu, V.~Taillandier, {Floating marine debris
  surface drift: Convergence and accumulation toward the South Pacific
  subtropical gyre}, Mar. Poll. Bull. 58 (2009) 1347--1355.

\bibitem{LawEtAl_2010}
K.~L. Law, S.~{Mor{\'e}t-Ferguson}, N.~A. Maximenko, G.~Proskurowski, E.~E.
  Peacock, J.~Hafner, C.~M. Reddy, {Plastic accumulation in the North Atlantic
  subtropical gyre}, Science 329 (2010) 1185--1188.

\bibitem{MaximenkoEtAl_2012}
N.~Maximenko, J.~Hafner, P.~Niiler, {Pathways of marine debris derived from
  trajectories of Lagrangian drifters}, Mar. Poll. Bull. 65 (2012) 51--62.

\bibitem{CozarEtAl_2014}
A.~Cozar, {et al}, {Plastic debris in the open ocean}, PNAS 111 (2014)
  10239--10244.

\bibitem{Kraichnan}
R.~H. Kraichnan, {Anomalous Scaling of a Randomly Advected Passive Scalar}, PRL
  72 (1994) 1016.

\bibitem{MaxeyRiley}
M.~R. Maxey, J.~J. Riley, {Equation of motion for a small rigid sphere in a
  nonuniform flow}, PoF 26 (1983) 883.

\bibitem{FalkovichI}
E.~Balkovsky, G.~Falkovich, A.~Fouxon, {Intermittent Distribution of Inertial
  Particles in Turbulent Flows}, PRL 86 (2001) 2790--2793.

\bibitem{QureshiEtAl_2007}
N.~M. Qureshi, M.~Bourgoin, C.~Baudet, A.~Cartellier, Y.~Gagne, {Turbulent
  Transport of Material Particles: An Experimental Study of Finite Size
  Effects}, PRL 99 (2007) 184502.

\bibitem{XuBodenchatz_2008}
H.~Xu, E.~Bodenschatz, {Motion of inertial particles with size larger than
  Kolmogorov scale in turbulent flows}, Phys. D 237~(14-17) (2008) 2095--2100.

\bibitem{SquiresEaton1991}
K.~D. Squires, J.~K. Eaton, {Preferential concentration of particles by
  turbulence}, PoF A 3 (1991) 1169.

\bibitem{Bourgoin}
R.~Monchaux, M.~Bourgoin, A.~Cartellier, {Preferential concentration of heavy
  particles: A Vorono{\"i} analysis}, PoF 22 (2010) 103304.

\bibitem{BourgoinXu2014}
M.~Bourgoin, H.~Xu, {Focus on dynamics of particles in turbulence}, New J.
  Phys. 16 (2014) 085010.

\bibitem{CressmanGoldburg_JSP2003}
J.~R. Cressman, W.~I. Goldburg, {Compressible Flow: Turbulence at the surface},
  J. Stat. Phys. 113 (2003) 875--883.

\bibitem{BoffettaEtAl_2004}
G.~Boffetta, J.~Davoudi, B.~Eckhardt, J.~Schumacher, {Lagrangian Tracers on a
  Surface Flow: The Role of Time Correlations}, PRL 93 (2004) 134501.

\bibitem{Goldburg}
J.~R. Cressman, J.~Davoudi, W.~I. Goldburg, J.~Schumacher, {Eulerian and
  Lagrangian studies in surface flow turbulence}, New J. Phys. 6 (2004) 53.

\bibitem{GoldburgPhysicaD}
J.~Larkin, W.~Goldburg, M.~M. Bandi, {Time evolution of a fractal distribution:
  particle concentrations in free-surface turbulence}, Phys. D 239 (2010)
  1264--1268.

\bibitem{Lovecchio}
S.~Lovecchio, C.~Marchioli, A.~Soldati, {Time persistence of floating-particle
  clusters in free-surface turbulence}, PRE 88 (2013) 033003.

\bibitem{HerterichHasselmann1982}
K.~Herterich, K.~Hasselmann, {The horizontal diffusion of tracers by surface
  waves}, JPO 12 (1982) 704--711.

\bibitem{VandenBroeck1999}
C.~{Van den Broeck}, {Stokes' drift: An exact result}, EPL 46 (1999) 1.

\bibitem{SantamariaEtAl2013}
F.~Santamaria, G.~Boffetta, M.~{Martins Afonso}, A.~Mazzino, M.~Onorato,
  D.~Pugliese, {Stokes drift for inertial particles transported by water
  waves}, EPL 102 (2013) 14003.

\bibitem{Falcovich_2005}
G.~Falkovich, A.~Weinberg, P.~Denissenko, S.~Lukaschuk, {Floater clustering in
  a standing wave}, Nature 435 (2005) 1045--1046.

\bibitem{SanliEtAl_2014a}
C.~Sanl\i, D.~Lohse, D.~{van der Meer}, {From antinode clusters to node
  clusters: The concentration-dependent transition of floaters on a standing
  Faraday wave}, PRE 89 (2014) 053011.

\bibitem{SanliEtAl_2014b}
C.~Sanl\i, K.~Saitoh, S.~Luding, D.~{van der Meer}, {Collective motion of
  macroscopic spheres floating on capillary ripples: Dynamic heterogeneity and
  dynamic criticality}, PRE 90 (2014) 033018.

\bibitem{Batchelor}
G.~K. Batchelor, {Introduction to fluid dynamics}, Cambridge University Press,
  1967.

\bibitem{Douady_1990}
S.~Douady, {Experimental study of the Faraday instability}, JFM 221 (1990)
  383--409.

\bibitem{GordilloMujica_2014}
L.~Gordillo, N.~Mujica, {Measurement of the velocity field in parametrically
  excited solitary waves}, JFM 754 (2014) 590--604.

\bibitem{Bondarenko}
N.~F. Bondarenko, M.~Z. Gak, F.~V. Dolzhanskiy, {Laboratory and theoretical
  models of plane periodic flow}, Izvestiya, Atmospheric and Oceanic Physics 15
  (1979) 711--716.

\bibitem{Sommeria_1986}
J.~Sommeria, {Experimental study of the two-dimensional inverse energy cascade
  in a square box}, JFM 170 (1986) 139--168.

\bibitem{Davidson}
P.~A. Davidson, {Magnetohydrodynamics in materials processing}, ARFM 31 (1999)
  273.

\bibitem{TabelingEtAl_1991}
P.~Tabeling, S.~Burkhart, O.~Cardoso, H.~Willaime, {Experimental study of
  freely decaying two-dimensional turbulence}, PRL 67 (1991) 3772--3775.

\bibitem{RiveraWuPRL2000}
M.~Rivera, X.~L. Wu, {External Dissipation in Driven Two-Dimensional
  Turbulence}, PRL 85 (2000) 976.

\bibitem{FauveTabeling}
P.~Tabeling, B.~Perrin, S.~Fauve, {Instability of a linear array of forced
  vortices}, EPL 3 (1987) 459--465.

\bibitem{Dauxois}
T.~Dauxois, {Nonlinear stability of counter-rotating vortices}, PoF 6 (1994)
  1625--1627.

\bibitem{Thess}
A.~Thess, {Instabilities in two-dimensional spatially periodic flows. Part I:
  Kolmogorov flow}, PoF A (1991) 1--11.

\bibitem{GollubMixing}
G.~A. Voth, G.~Haller, J.~P. Gollub, {Experimental Measurements of Stretching
  Fields in Fluid Mixing}, PRL 88 (2002) 254501.

\bibitem{FalconFauve_2009}
C.~Falc{\'o}n, S.~Fauve, {Wave-vortex interaction}, PRE 80 (2009) 056213.

\bibitem{GutierrezAumaitre_2016}
P.~Guti{\'e}rrez, S.~Auma{\^\i}tre, {Surface waves propagating on a turbulent
  flow}, PoF 28 (2016) 025107.

\bibitem{OuelletteTrackingLink}
N.~T. Ouellette, {Programs available on\\
  $http://web.stanford.edu/~nto/software-tracking.shtml$}.

\bibitem{OuelletteXuBodenchatz_2006}
N.~T. Ouellette, H.~Xu, E.~Bodenschatz, {A quantitative study of
  three-dimensional Lagrangian particle tracking algorithms}, Exp. Fluids 40
  (2006) 301--313.

\bibitem{Young}
Y.-K. Tsang, W.~R. Young, {Forced-dissipative two-dimensional turbulence: A
  scaling regime controlled by drag}, PRE 79 (2009) 045308(R).

\bibitem{GouillardPhD}
E.~Gouillard, {Etude de l'advection chaotique dans des m\'elangeurs \`a tiges,
  en \'ecoulements ouverts et ferm\'es}, Ph.D. thesis, Universit\'e Paris 6 -
  Pierre et Marie Curie (2008).

\bibitem{Bonamy}
D.~Bonamy, F.~Daviaud, L.~Laurent, M.~Bonetti, J.~P. Bouchaud, {Multiscale
  Clustering in Granular Surface Flows}, PRL 89 (2002) 034301.

\bibitem{Aste}
T.~Aste, T.~{Di Matteo}, {Emergence of Gamma distributions in granular
  materials and packing models}, PRE 77 (2008) 021309.

\bibitem{pumir}
P.~Castiglione, A.~Pumir, {Evolution of triangles in a two-dimensional
  turbulent flow}, PRE 64 (2001) 056303.

\bibitem{Maxey_JFM1987}
M.~R. Maxey, {The gravitational settling of aerosol particles in homogeneous
  turbulence and random flow fields}, JFM 174 (1987) 441--465.

\bibitem{DouadyCadotVortex1}
S.~Douady, Y.~Couder, M.~E. Brachet, {Direct observation of the intermittency
  of intense vorticity filaments in turbulence}, PRL 67 (1991) 983.

\bibitem{DouadyCadotVortex2}
O.~Cadot, S.~Douady, Y.~Couder, {Characterization of the low-pressure filaments
  in a three-dimensional turbulent shear flow}, PoF 7 (1995) 630.

\bibitem{Kamp}
L.~P.~J. Kamp, {Strain-vorticity induced secondary motion in shallow flows},
  PoF 24 (2013) 023601.

\bibitem{LovecchioEtAl_2015}
S.~Lovecchio, F.~Zonta, A.~Soldati, {Upscale energy transfer and flow topology
  in free-surface turbulence}, PRE 91 (2015) 033010.

\bibitem{MahadevanAmJPhys}
D.~Vella, L.~Mahadevan, {The ``Cheerios effect''}, Am. J. Phys. 73 (2005)
  817--825.

\bibitem{berhanu}
M.~J. Dalbe, D.~Cosic, M.~Berhanu, A.~Kudrolli, {Aggregation of frictional
  particles due to capillary attraction}, PRE 83 (2011) 051403.

\bibitem{vergassola}
A.~Celani, A.~Lanotte, A.~Mazzino, M.~Vergassola, {Universality and Saturation
  of Intermittency in Passive Scalar Turbulence}, PRL 84 (2000) 2385.

\bibitem{pumirPRL}
A.~Pumir, B.~I. Shraiman, M.~Chertkov, {Geometry of Lagrangian Dispersion in
  Turbulence}, PRL 85 (2000) 5324--5327.

\bibitem{ouellettePRL}
S.~T. Merrifield, D.~H. Kelley, N.~T. Ouellette, {Scale-Dependent Statistical
  Geometry in Two-Dimensional Flow}, PRL 104 (2010) 254501.

\end{thebibliography}

\end{document}